\documentclass[12pt]{article}

\usepackage{graphicx}

\begin {document}

\begin{center}

{\large \textbf{Single Heavy Flavour Baryons using Coulomb plus Power law
interquark Potential}} \\
\textbf{ Ajay Majethiya, Bhavin Patel and P. C. Vinodkumar}\\
Department of Physics, Sardar Patel University,
Vallabh Vidyanagar- 388 120, Gujarat, INDIA
\\
\end{center}

 \abstract{ Properties of single heavy flavor baryons in a non
relativistic potential model with colour coulomb plus power law
confinement potential have been studied. The ground state
 masses of single heavy baryons and the mass difference between the
($J^{P}=\frac{3} {2}^{+}$ and $J^{P}=\frac{1} {2}^{+}$) states
are computed using a spin dependent two body potential. Using the
spin-flavour structure of the constituting quarks and by defining
an effective confined mass of the constituent quarks within the
baryons, the magnetic moments are computed. The masses and
magnetic moments of the single heavy baryons  are found to be in
accordance with the existing  experimental values and with other
theoretical predictions. It is found that an additional attractive
interaction of the order of $-200$ Me$V$ is required for the
antisymmetric states of $\Lambda_{Q}$ (Q$\in c,b)$. It is also
found that the spin hyperfine interaction parameters play decisive
role in hadron spectroscopy.}

\section{Introduction}
Confirmation of the existence of the charmed baryons at Fermilab\cite{Baltay1979} arose
an increasing interest on heavy-baryon spectroscopy. It is striking that baryons
containing one or two heavy charm or beauty flavour could play an important role in our
understanding of QCD at the hadronic scale \cite{Garcilazo2007}. The copious production
of heavy quarks at LEP, Fermilab Tevatron, CERN and B factories, opened up rich
spectroscopic study of heavy hadrons. Many theoretical models
\cite{Roberts2007,Hwang2008,Amand2006,Giannini2001,Ebert2005,Yu2006,Bhavin2008,B.Silvestre-brac1996,Santopinto1998,Rosner609}
have also predicted the heavy baryon mass spectrum. The non-relativistic quark model
(NRQM) \cite{A.Valcarce2005} has been able to explain very nicely the mass spectrum of
light baryons. Though the experimental and theoretical data on the properties of heavy
flavour mesons are available plenty in literature, the masses of most of the heavy
baryons have not been measured yet experimentally \cite{arXiv:hep2000}. Thus the recent
predictions about the heavy baryon mass spectrum have become a subject of renewed
interest due to the experimental facilities at Belle, BABAR, DELPHI, CLEO, CDF etc
\cite{Mizuk2005,Aubert2007,Feindt2007,Edwards1995,Artuso2001,I.V.Gorelov2007}. These
experimental groups have been successful in discovering heavy baryonic states along
with other heavy flavour mesonic states and it is expected that more heavy flavour
baryon states will be detected in near future. Most of the new states are within the
heavy flavour sector with one or more heavy flavour content and some of them are far
from most of the theoretical predictions. Though there are consensus among the
theoretical predictions on the ground state masses \cite {Giannini2001,Ebert2005},
there are little agreement among the model predictions of the properties like
spin-hyperfine spiltting among the $J^{P}=\frac{1} {2}^{+}$ and $J^{P}=\frac{3}
{2}^{+}$ baryons, the form factors \cite{Giannini2001}, magnetic moments etc
\cite{Bhavin2008}. All these reasons make the study of the heavy flavour spectroscopy
extremely rich and interesting. The study of heavy baryons further provide excellent
laboratory to understand the dynamics of light
quarks in the vicinity of heavy flavour quarks in bound states.\\
For the present study of the properties of the low-lying baryonic states with a heavy
quark, we consider coulomb plus power form as the interquark interaction potential
\cite{Bhavin2008,Ajay2007,Bhavin20088}. The spin hyperfine interactions similar to the
one employed in \cite{Garcilazo2007} have been used with mass dependance of the
constituting quarks in the present study. The magnetic moments of the baryons are
computed based on the non-relativistic quark model using spin-flavour wave functions of
the constituting quarks \cite{Bhavin2008}.\\
\\The paper is organized as follows. In sect.II of this paper, the basic methodology adopted for the present
study of computing the binding energy of the constituting quarks within the baryon
containing single heavy quarks is described. In sect.III, we present the calculation of
magnetic moments of the baryons. In sect.IV, we present our results and discuss the
important features and conclusions of the present study.

\section{Methodology and Binding energy of the Single Heavy Flavour Baryons}
We start with the color singlet Hamiltonian of the system  as
\begin{equation}\label{eq:2.01}
H=-\sum\limits_{i=1}^3\frac{\nabla_{i}^2}{2m_{i}}+\sum \limits_{i<j} V_{ij}
\end{equation}
 Where, the interquark potential\begin{center}$V_{ij}$=$ -\frac{2 \alpha_{s}}{3}\frac{1}{x_{ij}} +\beta \ x_{ij}^\nu
+V_{spin}(ij)$;\end{center}
\begin{equation}
x_{ij}=|\overrightarrow{x_{i}}-\overrightarrow{x_{j}}|
\end{equation}
 Following ref \cite{Yu2006}, the
single particle position co-ordinates are replaced by the CM
co-ordinates plus the interquark distances of  $q_{1}$, $q_{2}$
and $q_{3}=Q$, as
\begin{eqnarray}
\overrightarrow{X}&=&\frac{1}{\sum{m_{i}}}\sum\limits_{i=1}^3{m_{i}}\overrightarrow{x_{i}};\\
\cr \overrightarrow{r_1}&=&\overrightarrow{x_1}-\overrightarrow{x_3}; \cr
\overrightarrow{r_2}&=&\overrightarrow{x_2}-\overrightarrow{x_3}\
\end{eqnarray}
and
\begin{eqnarray}
{r_{12}}&=&|\overrightarrow r_{2}- \overrightarrow r_{1}|= |\overrightarrow x_{2}-
\overrightarrow x_{1}| ;\\\cr\ m_{i3}&=&\frac{m_{i}m_{3}}{m_{i}+m_{3}};\ \ m_{3}=m_{Q}
\end{eqnarray}

\begin{table} \caption{The Model Parameters with the Variational Parameter $\lambda$} \vspace{0.01in}
\begin{center}
\label{tab:01}
\begin{tabular}{llll}
\hline System&$\nu$& $\lambda$ in units of & $\beta$ in units of\\
&&Bohr radius&(Bohr Energy)$^{\nu+1}$\\ \hline
Charm Baryons&0&1.31&1.018\\
  &1&3.07&1.262\\
 &2&3.64&1.830\\
\hline
Beauty Baryons &0&1.31&4.782\\
 &1&5.57&10.000\\
 &2&6.72&26.530\\
  \hline
\end{tabular}
\end{center}
$m_{u}=m_{d}=338,$ $m_{s} =420,$ $m_{c}
=1380,$ $m_{b}=4275$ (in Me$V$)\\
\end{table}

\begin{table}
\begin{center}
\caption{Hyperfine Parameters for the symmetric spin-flavour combinations of single
heavy baryons} \vspace{0.001in} \label{tab:02}

\begin{tabular}{@{}llll}
\hline
&\multicolumn{3}{l}{\textbf{\underline{$A$ with $(r_{0}=10^{-6}$ Me$V^{-1})$}}}\\
 System&$\nu=0$ &$\nu=1$ &$\nu=2$\\
\hline
cqq&13&1 &0.625\\
cqs&39&3 &1.875\\
bqq&78&1 &0.625\\
bqs&468&6& 3.750\\
\hline

\end{tabular}
\end{center}
\end{table}

Accordingly, the Hamiltonian can be separated into the
translationally invariant CM part and the part governing the
relative motion given by, $h$ as
\begin{equation}
h=-\frac{\nabla_{r1}^{2}}{2m_{13}}-\frac{\nabla_{r2}^{2}}{2m_{23}}
-\frac{\nabla_{r1}\cdot\nabla_{r2}}{m_3}+ V(r_{1},r_{2})
\end{equation}
 Here, $m_{1},m_{2},m_{3}$ are the constituent quark masses and all the independent orientations
of co-ordinates $\overrightarrow{r_1}$ and $\overrightarrow{r_2}$
are assumed. Under the spherically symmetric approximation, the
potential is then written in terms of the magnitudes of the
co-ordinates $\overrightarrow{r_1}$ and $\overrightarrow{r_2}$ as \\
\begin{equation}\label{eq:2.8}
V(r_{1},r_{2})=-\frac{2\alpha_{s}}{3}
\left(\frac{1}{r_{1}}+\frac{1}{r_{2}}+\frac{1}{r_{12}}\right) +
\beta (r_{1}^\nu+r_{2}^\nu+r_{12}^\nu)+V_{spin}\\
\end{equation}
Here, $\alpha_{s}$ is the running strong coupling constant and $V_{spin}$ is the spin
dependent part of the three body system. We choose the spatial part of the trial wave
function as \cite{Yu2006}
\begin{equation}
\Psi(r_{1},r_{2})=f(r_{1})f(r_{2})
\end{equation}
with a normalization condition given by\\
 $
\int|\Psi(r_{1},r_{2})|^{2}d^{3}r_{1}d^{3}r_{2}=1$.\\

Here, $f(r)=\frac{\lambda^{3/2}}{\sqrt{\pi}}\ e^{-\lambda{r}}$ and
$\lambda$ is a common variational parameter corresponds to both $f(r_{1})$ and $f(r_{2}).$\\
We find the expression for the ground state energy E without the spin dependent part of
the potential given by Eqn \ref{eq:2.8}, as

\begin{eqnarray}\label{eq:2.10}
   E &=&-\lambda^2+2\lambda(\lambda-1)-\frac{5\lambda}{8}+
8\beta\lambda^{3}2^{-3-\nu}\lambda^{-3-\nu}\Gamma(\nu+3)
 \cr &&+ \frac{8 \beta
2^{-\nu}\Gamma(\nu+6)G_{\nu}}{\nu+2} \lambda^{-\nu}
\end{eqnarray}

\begin{figure}
\begin{center}
\resizebox{0.5 \textwidth}{!}{%
\includegraphics{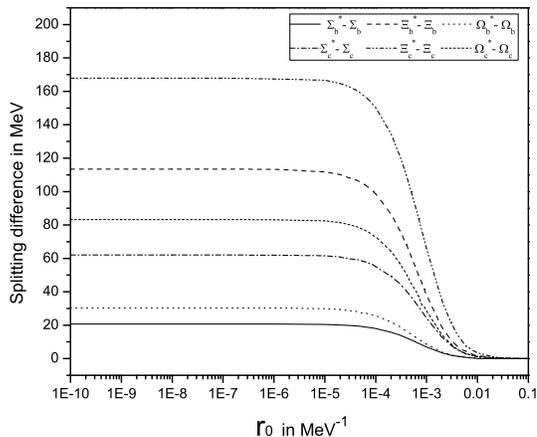}}
\caption{Spiltting  difference for symmetric states vs $r_{0}$.}\label{fig:01}
\end{center}
\end{figure}

\begin{figure}
\begin{center}
\resizebox{0.5 \textwidth}{!}{%
\includegraphics{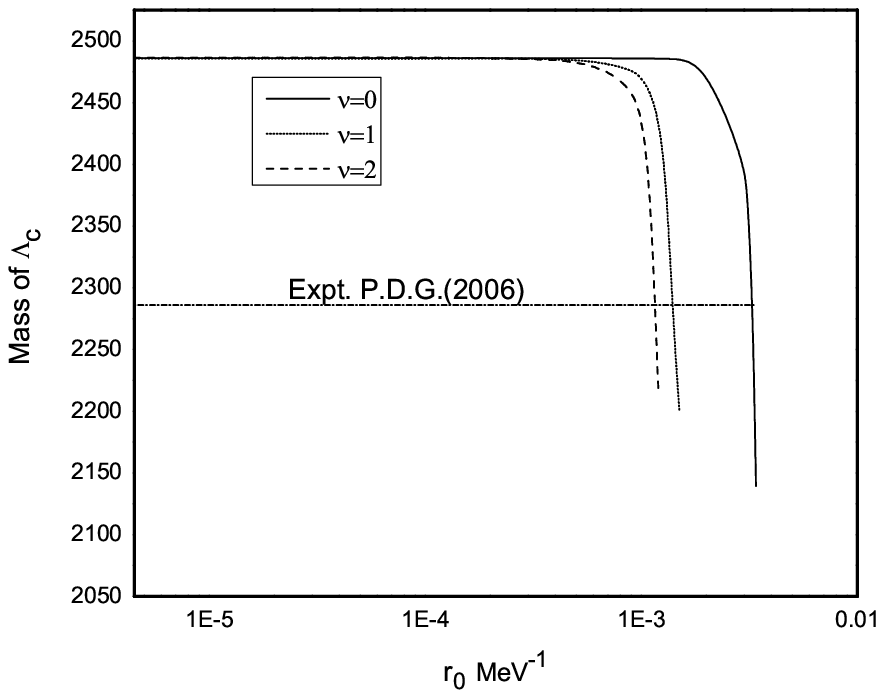}}
\caption{Dependance of $r_{0}$ on the masses  of $\Lambda_{c}$.} \label{fig:02}
\end{center}
\end{figure}

\begin{figure}
\begin{center}
\resizebox{0.5 \textwidth}{!}{%
\includegraphics{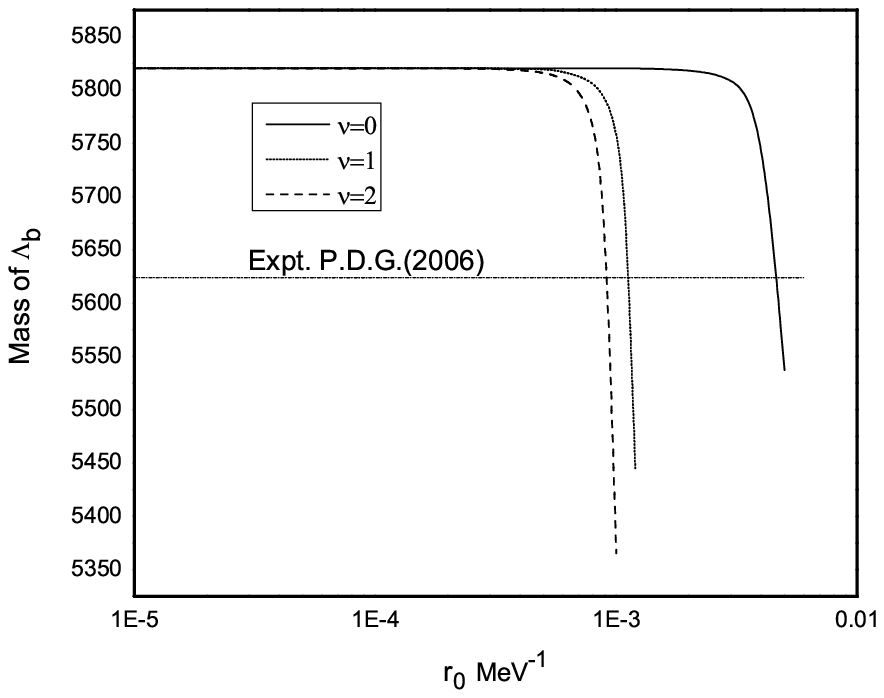}}
\caption{Dependance of $r_{0}$ on the masses of $\Lambda_{b}$.}\label{fig:03}
\end{center}
\end{figure}

Where,\\
$G_{\nu}$=$\frac{1}{128}$[1-$(-1)^{2\nu+6}$\,\,$_2$F$_1$(1;\,$\nu$+6;\,3;\,2)]
        -
$\frac{1}{192}$[1+$(-1)^{2\nu+6}$\,\,$_2$F$_1$(1;\,$\nu$+6;\,4;\,2)]\\  \\ and
$_2$F$_1$ is the Hypergeometric function. We minimize the energy expression
given by  Eqn \ref{eq:2.10} to find the variational parameter $\lambda$. \\
In the single heavy quark baryonic system,  the heavy quark acts as a static colour
source, with two light quarks revolving around. For simplicity, we will take the
baryonic units, in which all the length and energy scales are measured in the unit of
the Bohr radius $(m_{red} 2  \alpha_s/3 )^{-1}$ and Bohr energy $ m_{red}( 2 \alpha_s/3
)^{2}$ as in \cite{Yu2006}. Where $ m_{red}$ is computed with reference to the
combination of the lightest flavour with the heavy flavour at the center, as $\
m_{lQ}=\frac{m_{l}m_{Q}}{m_{l}+m_{Q}}$, $m_{l}$ is mass of the lightest flavour quark.
For example, in the case of qqc and qsc, we consider
$m_{red}=\frac{m_{q}m_{Q}}{m_{q}+m_{Q}}$, while in case of the ssc,
$m_{red}=\frac{m_{s}m_{Q}}{m_{s}+m_{Q}}$. The running strong coupling constant is
computed using the relation
\begin{equation}
\alpha_s(\mu)=\frac{\alpha_s(\mu_0)}{1+\frac{33-2\,n_f}{12\,\pi}\alpha_s(\mu_0)ln(\frac{\mu}{\mu_0})}
\end{equation}
For the present study, we considered $\alpha_s(\mu_{0}=1 GeV)\approx 0.7$. Though it is
an ad-hoc choice, the same value has been employed in our earlier studies on
light-heavy mesons \cite{Ajay2005}. The energy eigen value given by
 Eqn \ref{eq:2.10} is computed for the choices of the potential index
$\nu = 0, 1$ and $2$ respectively. The spin average mass of baryonic system (without
spin contribution) is then obtained as
\begin{equation}\label{eq:2.12}
M_{Qqq}=\Sigma {m}_{i}+E\end{equation} The quark masses and the potential parameters of
the model as listed in Table \ref{tab:01} are obtained so as to get the ground state
spin average masses of qqc (2486 Me$V$), qsc (2568 Me$V$), ssc (2730 Me$V$),
qqb (5820 Me$V$), qsb (5902 Me$V$) and ssb (6176 Me$V$) for each potential index $\nu=$0, 1 and 2.\\
One of the most tricky and important components of hadron spectroscopy is to predict
correctly the spin hyperfine split among the $J^{P}=\frac{3}{2}^{+}$ and
$J^{P}=\frac{1}{2}^{+}$ baryons. There exist many attempts starting from the standard
OGE potential corresponds to a free gluon exchange to confined gluon exchange
\cite{Vijayakumar1993,Vinodkumar1999,Vinodkumar1992}. The free gluon exchange
interaction between the point like quarks leads to the delta function behavior, while
the effect of confinement of the gluons as well as the finite size of the quarks
warrant smoothening of the delta function. One of the most suitable form is provided by
\cite{Garcilazo2007}
\begin{equation} \label{eq:2.13}
V_{spin}=
-\frac{A}{4}\alpha_s\sum\limits_{i<j}\overrightarrow{\lambda_{i}}.\overrightarrow{\lambda_{j}}
\frac{ e^{\frac{-r_{ij}}{r_{0}}}}{r_{ij}r_{0}^2}
\frac{\overrightarrow{\sigma_{i}}.\overrightarrow{\sigma_{j}}}{6m_{i}m_{j}}
\end{equation}\\
Here, the parameter $A$ and the regularization parameter $r_{0}$ are considered as the
hyperfine parameters of the model. It is closely similar to the form given by
\cite{Garcilazo2007,Bhavin2008} except the way we treat the parameter $r_{0}$. Here we
treat $r_{0}$ as a hyperfine parameter related to gluon dynamics, and hence independent
of the masses of the interacting quarks as treated by \cite{Garcilazo2007}. As no well
established procedure to evaluate $r_{0}$ from colour gluon dynamics is known, we
obtain the optimum values of these hyperfine parameters graphically by studying the
behavior of the hyperfine split of $\Sigma^{*}_{Q}-\Sigma_{Q}$, $\Xi^{*}_{Q}- \Xi_{Q}$
and $\Omega^{*}_{Q}-\Omega_{Q}$ for both the charm ($Q = c$) and beauty ($Q = b$)
baryons at different $r_{0}$ values. The behavior is shown in Fig \ref{fig:01}. It is
seen that below the range of $r_{0}<10^{-4}\ $Me$V^{-1}$ as seen in Fig \ref{fig:01},
the split gets saturated while for $r_{0} \geq 0.01$, the corresponding hyperfine mass
difference approaches to zero. Thus we fix our regularization parameter well within the
saturation region of $r_{0}$ value to about $10^{-6}$ Me$V^{-1}$. The other hyperfine
parameter, $A$ of Eqn \ref{eq:2.13} for different choices of the power index $\nu$ and
for the different quark combinations (Qqq) are listed in Table \ref{tab:02}. The
computed masses of $J^{P}=\frac{1} {2}^{+}$ and $J^{P}=\frac{3} {2}^{+}$ of the
symmetric compositions of Qqq states are listed in Table \ref{tab:03}
along with other contemporary model predictions and with known experimental values. \\
\\
In the case of spin
antisymmetric state of $\Lambda_{Q}$ baryons, similar study on the masses of
$\Lambda_{c}$ and $\Lambda_{b}$ with $r_{0}$ is shown in Fig \ref{fig:02} and
\ref{fig:03} respectively. A similar saturation property of the $\Lambda_{Q}$ masses
with $r_{0}$ below $10^{-4}$Me$V^{-1}$ is observed. However, the saturated masses with
respect to $r_{0}$ of $\Lambda_{b}$(5820 Me$V$) and $\Lambda_{c}$(2486 Me$V$) as seen
from Fig \ref{fig:02} and \ref{fig:03} are about 200 Me$V$ more than their experimental
masses of 5624 Me$V$ and 2286 Me$V$ respectively. It suggests then the requirement of
an attractive interaction of about 200 Me$V$ magnitude to the antisymmetric state
irrespective of its heavy quark content and potential index. One can add such
attractive part to the Hamiltonian of the $\Lambda_{Q}$ state or can absorb it in the
regularization parameter of the spin-hyperfine interaction. Hence, we attain $r_{0}$
values away from the mass saturation region nearer to the pole, as shown in Fig
\ref{fig:02} and \ref{fig:03}. Accordingly, without changing $A$ parameters for
$\Lambda_{Q}(Q\in c,b)$, we get the regularization parameter $r_{0}$ corresponding to
the experimental masses of $\Lambda_{c}$ and $\Lambda_{b}$. The details are summarized
in Table \ref{tab:04}. Our results are also compared with other model predictions as
well as with experimental values. The mass difference between
$\Sigma^{*}_{Q}-\Lambda_{Q}$, $\Xi^{*}_{Q}-\Xi_{Q}$ and $\Sigma^{*}_{Q}-\Sigma_{Q}$ are
also listed in Table \ref{tab:05} along with the experimental values as well as with
other model predictions.

\begin{table*}
\begin{center}
\caption{Single heavy baryon masses of symmetric spin-flavour combinations (masses are
in MeV)} \vspace{0.001in} \label{tab:03}
\begin{tabular}{@{}lllllll}\\
\hline Baryon &Quark Content&$\nu$
&{$\textbf{J}^P=\frac{1}{2}^+$}&Others&{$\textbf{J}^P=\frac{3}{2}^+$}&Others\\
\hline
$\Sigma_{c}$ &qqc&$0 $&2444&2453{\cite{Amand2006}}&2507&$-$\\
&& $1$ &2445& 2451{\cite{Bhavin2008}}&2507&2516{\cite{Bhavin2008}} \\
&& $2$ &2443&2460$\pm$80{\cite{Bowler1998}}&2508&2440$\pm$70{\cite{Bowler1998}} \\
&&\ \ \ \ \ \ &&2454{\cite{PDG2006}} && 2518{\cite{PDG2006}}\\
&&\ \ \ \ \ \ &&2448{\cite{Garcilazo2007}} && 2505{\cite{Garcilazo2007}}\\
\hline

 $\Xi_{c} $&qsc&$0 $&2455&2466{\cite{Amand2006}}& 2625&$-$\\
&&$1$ &2464&2485{\cite{Bhavin2008}}& 2637&2672{\cite{Bhavin2008}}\\
&&$2$ &2452&2468{\cite{Roncaglia1995}}&2627 &2650{\cite{Roncaglia1995}}\\
&&\ \ \ \ \ \ &&2473{\cite{Mathur2002}}&&2680{\cite{Mathur2002}}\\
&&\ \ \ \ \ \ &&2496{\cite{Garcilazo2007}}&&2633{\cite{Garcilazo2007}}\\
&&\ \ \ \ \ \ &&2468{\cite{PDG2006}}&&2646{\cite{PDG2006}}\\
&&\ \ \ \ \ \ && 2481{\cite{Ebert2005}} && 2654{\cite{Ebert2005}} \\
\hline $\Omega_{c}

$&ssc&$0 $&2674&2698{\cite{Amand2006}}& 2758&$-$\\
&&$1$&2674&2696{\cite{Bhavin2008}}& 2758& 2757{\cite{Bhavin2008}}\\
&&$2$&2673&2710{\cite{Roncaglia1995}}&2759&2770{\cite{Roncaglia1995}}\\
&&\ \ \ \ \ \ &&2678{\cite{Mathur2002}}&&2752{\cite{Mathur2002}}\\
&&\ \ \ \ \ \ &&2701{\cite{Garcilazo2007}}&&2759{\cite{Garcilazo2007}}\\
&&\ \ \ \ \ \ &&2698{\cite{Ebert2005}} &&2768{\cite{Ebert2005}} \\
\hline
$\Sigma_{b}$&qqb&$0$&5806&5820{\cite{Amand2006}}&5827&$-$\\
&&$1$ &5807&5801{\cite{Bhavin2008}}&5827& 5823{\cite{Bhavin2008}} \\
&&$2$ &5805& 5806$\pm$70{\cite{Bowler1998}}&5828& 5780$\pm$70{\cite{Bowler1998}}  \\
&&\ \ \ \ \ \ &&5805{\cite{Ebert2005}}&& 5834{\cite{Ebert2005}}\\
&&\ \ \ \ \ \ &&5808{\cite{I.V.Gorelov2007}}&& 5829{\cite{I.V.Gorelov2007}}\\
&&\ \ \ \ \ \ &&5789{\cite{Garcilazo2007}}&&5844{\cite{Garcilazo2007}}\\

\hline
 $\Xi_{b} $&qsb&$0$&5826&5624{\cite{Amand2006}}&5940&$-$\\
&&$1$&5827&5872{\cite{Bhavin2008}}&5939&5936{\cite{Bhavin2008}}\\
&&$2$ &5820&5820{\cite{Roncaglia1995}}&5943&5980{\cite{Roncaglia1995}}\\
&&\ \ \ \ \ \ &&5847{\cite{Mathur2002}}&&5959{\cite{Mathur2002}}\\
&&\ \ \ \ \ \ &&5825{\cite{Garcilazo2007}}&&5967{\cite{Garcilazo2007}}\\
&&\ \ \ \ \ \ &&5805{\cite{Ebert2005}} &&5963{\cite{Ebert2005}} \\

\hline
 $\Omega_{b} $&ssb&$0 $&6156&6040{\cite{Amand2006}}&6187&$-$\\
&&1 &6156 &6005{\cite{Bhavin2008}}& 6186& 6065{\cite{Bhavin2008}}\\
&&2 &6154&6060{\cite{Roncaglia1995}}&6187&6090{\cite{Roncaglia1995}}\\
&&\ \ \ \ \ \ & &6040{\cite{Mathur2002}}&&6060{\cite{Mathur2002}}\\
&&\ \ \ \ \ \ & &6037{\cite{Garcilazo2007}}&&6090{\cite{Garcilazo2007}}\\
&&\ \ \ \ \ \ &&6065{\cite{Ebert2005}}&& 6088{\cite{Ebert2005}}\\

\hline

\end{tabular}
\end{center}
 \end{table*}

\begin{table*}[t]
\begin{center}
\caption{The Hyperfine Parameters for $\Lambda_{Q}$ Baryons}\vspace{0.001in}
\label{tab:04}
\begin{tabular}{@{}lllll}
\hline Baryon ($\Lambda_{Q}$) &$\nu$&$A$ &$r_{0}$ in Me$V^{-1}$\\
\hline
 $\Lambda_{c} (2286) $&$0$&13&$3.266\times10^{-3}$& \\
&$1$&1&$1.395\times10^{-3}$& \\
&$2$&$0.625$&$1.173\times10^{-3}$&\\

\hline
 $\Lambda_{b}(5624) $&$0$&$78$&$4.860\times10^{-3}$&\\
&$1$&$1$&$1.144\times10^{-3}$& \\
&$2$&$0.625$&$0.940\times10^{-3}$&\\

\hline

\end{tabular}
\end{center}
\end{table*}

\begin{table*} [t]\caption{$J^{P}=\frac{3}{2}^+$ and $\frac{1}{2}^+$ ground state Mass
difference at different Potential index, $\nu$ (Masses in Me$V$)}
\begin{center}
\label{tab:05}
\begin{tabular}{@{}lllllll}

 &&&&\textbf{$$}&\\
\hline
Mass difference&$\nu=0$ &$\nu=1$&$\nu=2$&Expt.&\cite{Garcilazo2007}&\cite{Bhavin2008}\\
\hline
$\Sigma^{*+}_{c}-\Lambda^{+}_{c}$&221&221&222&232&233&235\\

$\Sigma^{*+}_{c}-\Sigma^{+}_{c}$&63&62&65&64&57&65\\

$\Sigma^{*0}_{b}-\Lambda^{0}_{b}$&203&203&204&205&220&205\\

$\Sigma^{*0}_{b}-\Sigma^{0}_{b}$&21&20&23&21&55&23\\

$\Xi^{*0}_{c}-\Xi^{0}_{c}$& 170&173&175&178&137&187\\

$\Xi^{*0}_{b}-\Xi^{0}_{b}$&114&112&123&$-$&178&64\\
\hline
\end{tabular}
\end{center}
\end{table*}

\begin{table*}
\begin{center}
\caption{Magnetic moments of single heavy charm and beauty baryons in terms of Nuclear
magneton $\mu_{N}$(* indicates $J^P=\frac{3}{2}^+$ state.)}\label{tab:06}
\begin{tabular}{@{}lcccccccc}
 &&&&\textbf{$$}&\\
\hline
Baryon& $\nu=0$&&$\nu=1$&$\nu=2$&\cite{Bhavin2008}&\cite{B.Silvestre-brac1996}&RQM\cite{Amand2006}&NRQM\cite{Amand2006}\\
\hline
$\Sigma^{++}_{c}$&1.9487&&1.9479&1.9494&2.2720&2.5320&                          1.7600&1.8600\\
$\Sigma^{*++}_{c}$&3.4071&&3.4071&3.4058&3.8420&$-$&                                    $-$&$-$\\
\hline
$\Sigma^{+}_{c}$&0.3918&&0.3917&0.3920&0.5000&0.5480&0.3600&0.3700\\
$\Sigma^{*+}_{c}$&1.1306   &&1.1306&1.1301&1.2520&$-$&$-$&$-$\\
\hline
$\Xi^{+}_{c}$&0.5105&&0.5087&0.5112&0.7090&0.2110&0.4100&0.3700\\
$\Xi^{*+}_{c}$&1.2700  &&1.2642&1.2690&1.513&$-$&$-$&$-$\\
\hline
$\Omega^{0}_{c}$&-0.9497&&-0.9497&-0.9501&-0.9580&-0.8350&-0.8500&-0.8500\\
$\Omega^{*0}_{c}$& -0.8339  &&-0.8339&-0.8336&-0.8650&$-$&$-$&$-$\\
\hline
$\Sigma^{0}_{c}$&-1.1650&&-1.1645&-1.1660&-1.0120&-1.4350&-1.0400&-1.1100\\
$\Sigma^{*0}_{c}$&-1.1460   &&-1.1460&-1.1455&-0.8480&$-$&$-$&$-$\\
\hline
$\Xi^{0}_{c} $&-1.1011&&-1.0971&-1.1025&-0.9640&0.3600&-0.9500&-0.9800\\
$\Xi^{*0}_{c} $&-0.9910   &&-0.9865&-0.9902&-0.6880&$-$&$-$&$-$\\
\hline
$\Sigma^{+}_{b}$&2.1249&&2.1246&2.1253&2.2260&2.6690&2.0700&2.0100\\
$\Sigma^{*+}_{b}$&3.0827   &&3.0827&3.0821&3.2390&$-$&$-$&$-$\\
\hline
$\Sigma^{0}_{b}$&0.54683&&0.54673&0.54692&0.5910&0.6820&0.5300&0.5200\\
$\Sigma^{*0}_{b}$&0.72404   &&0.72404&0.72392&0.7910&$-$&$-$&$-$\\
\hline
$\Omega^{-}_{b}$&-0.8047&&-0.8047&-0.8050&-0.9580&-0.7030&-0.8200&-0.7100\\
$\Omega^{*-}_{b}$& -1.2918  &&-1.2920&-1.2918&-1.1990&$-$&$-$&$-$\\
\hline
$\Sigma^{-}_{b}$&-1.0313&&-1.0311&-1.0315&-1.0450&-1.3050&-1.0100&-0.9700\\
$\Sigma^{*-}_{b}$&-1.6346  &&-1.6346&-1.6343&-1.6550&$-$&$-$&$-$\\
\hline
$\Xi^{0}_{b}$&0.6580&&0.6579&0.6587&0.7650& &0.6600&0.6500\\
$\Xi^{*0}_{b}$&0.8751  &&0.8753&0.8747&1.0410&$-$&$-$&$-$\\
\hline
$\Xi^{-}_{b}$&-0.9407&&-0.9406&-0.9417&-0.9010&-0.0550&-0.9100&-0.8400\\
$\Xi^{*-}_{b}$&-1.4770   &&-1.4773&-1.4762&-1.0950&$-$&$-$&$-$\\
\hline
$\Lambda^{+}_{c}$&0.4077&&0.4077&0.4077&0.3840&0.3410&0.3800&0.3700\\
$\Lambda^{0}_{b}$&-0.0640   &&-0.0640&-0.0640&-0.0640&-0.0600&$-0.0690$&$-0.0600$\\
\hline
\end{tabular}
\end{center}
\end{table*}

\begin{table*}
\begin{center}
\caption{ Baryon-meson mass Inequalities for $J^P=\frac{3}{2}^+$ single heavy Baryons.
(masses are in MeV)} \vspace{0.001in} \label{tab:07}
\begin{tabular}{@{}lllllll}\\
\hline  &Baryon-meson Inequalities&$\nu$
&Baryon mass && R.H.S of the inequality \\
  &&
&in Me$V$ &&  relation\cite{Shmuel Nussinov2002}\\
\hline  &$m_{\Sigma^{++}_{c}}(qqc)\geq\frac{1}{2}(m_{\rho}+2 m_{D^{*}})$&$0
$&2507&$\geq$&2394\\
&& $1$ &2507&$\geq$&\\
&& $2$ &2508& $\geq$&\\

\hline

&$m_{\Xi_{c}}(qsc)\geq\frac{1}{2}(m_{K^{*}}+ m_{D^{*}}+m_{D_{}s^{*}})$&$0 $&2625&$\geq$&2507\\
&&$1$ &2637&$\geq$\\
&&$2$ &2627&$\geq$\\

\hline&$m_{\Omega_{c}}(ssc)\geq\frac{1}{2}(m_{\phi}+2 m_{D^{*}})$&$0 $&2758&$\geq$&2518\\
&&$1$&2758&$\geq$\\
&&$2$&2759&$\geq$\\

\hline
&$m_{\Sigma_{b}}(qqb)\geq\frac{1}{2}(m_{\rho}+2 m_{B^{*}})$&$0$&5827&$\geq$&5710\\
&&$1$ &5827&$\geq$\\
&&$2$ &5828&$\geq$\\
\hline
&$m_{\Xi_{b}}(qsb)\geq\frac{1}{2}(m_{K^{*}}+ m_{B^{*}}+m_{B_{}s^{*}})$&$0$&5940&$\geq$&5793\\
&&$1$&5939&$\geq$\\
&&$2$ &5943&$\geq$\\

\hline
&$m_{\Omega_{b}}(ssb)\geq\frac{1}{2}(m_{\phi}+2 m_{B_{s}^{*}})$&$0 $&6187&$\geq$&5879\\
&&1 &6186 &$\geq$\\
&&2 &6187&$\geq$\\

\hline
\end{tabular}
\end{center}
 \end{table*}

\begin{table*}
\begin{center}
\caption{Baryon-meson mass Inequalities for $J^P=\frac{1}{2}^+$ single heavy Baryons.
(masses are in MeV)} \vspace{0.001in} \label{tab:08}
\begin{tabular}{@{}lllllll}\\
\hline  &Baryon-meson Inequalities&$\nu$
&Mass && R.H.S of the  \\
  &&
&in Me$V$ && inequa. relation\cite{Shmuel Nussinov2002}\\
\hline
&$m_{\Sigma_{c}}(qqc)\geq\frac{1}{4}(2m_{\rho}+3m_{D}+ m_{D^{*}})$&$0 $&2444&$\geq$&2288\\
&& $1$ &2445& $\geq$& \\
&& $2$ &2443&$\geq$& \\

\hline

&$m_{\Xi_{c}}(qsc)\geq\frac{1}{4}[(m_{K}+m_{K^{*}})+ (m_{D_{s}}+m_{D})+(m_{D^{*}}+m_{D_{}s^{*}})]$&$0 $&2455&$\geq$&2336\\
&&$1$ &2464&$\geq$\\
&&$2$ &2452&$\geq$\\

\hline &$m_{\Omega_{c}}(ssc)\geq\frac{1}{4}(2m_{\phi}+3m_{D_{s}}+ m_{D_{s}^{*}})$&$0 $&2674&$\geq$&2514\\
&&$1$&2674&$\geq$\\
&&$2$&2673&$\geq$\\

\hline
&$m_{\Sigma_{b}}(qqb)\geq\frac{1}{4}(2m_{\rho}+3m_{B}+ m_{B^{*}})$&$0$&5806&$\geq$&5675\\
&&$1$ &5807&$\geq$ \\
&&$2$ &5805& $\geq$ \\

\hline
&$m_{\Xi_{b}}(qsb)\geq\frac{1}{4}[(m_{K}+m_{K^{*}})+ (m_{B_{s}}+m_{B})+(m_{B^{*}}+m_{B_{}s^{*}})]$&$0$&5826&$\geq$&5682\\
&&$1$&5827&$\geq$\\
&&$2$ &5820&$\geq$\\

\hline
&$m_{\Omega_{b}}(ssb)\geq\frac{1}{2}(m_{\phi}+2 m_{B_{s}^{*}})$&$0 $&6156&$\geq$&$5880$\\
&&1 &6156 &$\geq$\\
&&2 &6154&$\geq$\\
\hline
&$m_{\Lambda_{c}}(qqc)\geq\frac{1}{4}(2m_{\pi}+3m_{D}+ m_{D^{*}})$&$0 $&2286&$\geq$&1973\\
&& $1$ &2286& $\geq$& \\
&& $2$ &2286&$\geq$& \\
\hline
&$m_{\Lambda_{b}}(qqb)\geq\frac{1}{4}(2m_{\pi}+3m_{B}+ m_{B^{*}})$&$0 $&5624&$\geq$&5359\\
&& $1$ &5624& $\geq$& \\
&& $2$ &5624&$\geq$& \\

\hline
\end{tabular}
\end{center}
 \end{table*}

\section{Magnetic Moments of the Single Heavy Flavour Baryons }
For the computation of the magnetic moments, we consider the mass of bound quarks
inside the baryons as its effective mass taking in to account of its binding
interactions with other two quarks described by the Hamiltonian given in Eqn
\ref{eq:2.01}. The effective mass for each of the constituting quark  $m^{eff}_{i}$ can
be defined as \cite{Bhavin2008}
\begin{equation}
m^{eff}_{i}=m_i\left(
1+\frac{\left<H\right>}{\sum\limits_{i}m_{i}}\right) \\
\end{equation}
such that the corresponding mass of the baryon is given by
\begin{equation}
M_B=\sum\limits_{i}m_{i}+\left<H\right>=\sum\limits_{i}m^{eff}_{i}\\
\end{equation}
Here, $\left<H\right>$ includes the spin hyperfine interaction also. For example, the
effective mass of the $u$ and $d$ quark from the chosen values of the mass parameter
will be different when it is in udc combinations
 or in udb combinations as $\frac{\left<H\right>_{udc}}{\sum\limits_{i}m_{i}}\neq\frac{\left<H\right>_{udb}}{\sum\limits_{i}m_{i}}$.
Now, the magnetic moment of baryons are obtained in terms of the bound quarks as
\begin{equation}
\mu_B=\sum\limits_{i}\left<\phi_{sf}\mid\mu_{i}
\overrightarrow{\sigma}_{i}\mid\phi_{sf}\right>
\end{equation}
where
\begin{equation}
\mu_{i}=\frac{e_{i}}{2m_{i}^{eff}}
\end{equation}
Here, $e_{i}$ and $\sigma_{i}$ represents the charge and the spin of the quark
constituting the baryonic state. We have employed the spin flavour wavefunction
($\left|\phi_{sf}\right>$) of the symmetric and antisymmetric states of the baryons as
used in \cite{Bhavin2008}. Using the spin flavour wave functions corresponds to
$J^{P}=\frac{1} {2}^{+}$ and $J^{P}=\frac{3} {2}^{+}$, we compute the magnetic moments
of the baryons containing a single charm or beauty quark. Our results are listed in
Table \ref{tab:06} for the choices of $\nu=0, 1$ and $2$. Other theoretical model
predictions of the magnetic moments are also listed for comparison.
\section{Results and Discussions} We have employed a simple nonrelativistic approach with coulomb plus
power potential to study the masses and magnetic moments of the single heavy flavour
baryons. The model parameters for each choices of the potential index, $\nu$ are listed
in Table \ref{tab:01} along with the corresponding variational parameter $\lambda$ of
the trial wavefunction. The model parameters are obtained to get the ground state spin
average masses of the Qqq systems.\\
\\ The spin hyperfine interaction has been taken into account according to
\cite{Garcilazo2007} where the interaction is expressed in terms of two hyperfine
parameters, $A$ and a regularization parameter $r_{0}$. It is interesting to see that
the hyperfine mass splitting of the symmetric spin combinations of the two body
interaction gets saturated with $r_{0}$ within a range, $r_{0}<10^{-4}$
 Me$V^{-1}$. Similar saturation property for the masses of $\Lambda_{Q}(Q\in b,c)$ is
also seen with respect $r_{0}$. The spin hyperfine parameter $A$,
 is now fixed to yield the experimental mass difference of $\Sigma^{*}_{Q}-\Sigma_{Q}$ well within the saturated range of $r_{0}=10^{-6}$
Me$V^{-1}$. The resulting hyperfine parameter, $A$ for different choices of $\nu$ are
shown in Table \ref{tab:02} for different quark compositions. It can be seen that the
parameter $A$ vary from $\nu=0$ to 2 according to a fixed ratio given by
$13:1:\frac{5}{8}$ for the cqq system and as $78:1:\frac{5}{8}$ in the case of bqq
systems. For the $\Xi_{Q}$ states, it is found that the ratio triples in the case of
$\Xi_{c}$ as $\nu$ varies from $0$ to $2$ and becomes six times in the case of
$\Xi_{b}$. It probably corresponds to a statistical factor related to the combination
of the light quark composition $3C_{2}$ with charm (cqs) and $4C_{2}$ in the case of
(bqs) systems.\\
\\  In Table \ref{tab:05}, we provide the $\Sigma^{*}_{Q}-\Lambda_{Q}$,
$\Xi^{*}_{Q}-\Xi_{Q}$ and $\Sigma^{*}_{Q}-\Sigma_{Q}$ mass differences obtained from
the present study for each case of the model potential, $\nu=0$ to $2$. The masses of
$J^{P}=\frac{1} {2}^{+}$ and $J^{P}=\frac{3} {2}^{+}$ baryons as listed in Table
\ref{tab:03} are in agreement with the well known baryon-meson mass inequalities
\cite{Shmuel Nussinov2002}. The inequality relations are shown in Table \ref{tab:07}
and \ref{tab:08} for the $J^{P}=\frac{3} {2}^{+}$ and $\frac{1} {2}^{+}$ states
respectively. The masses of single heavy baryons studied here are found to be well
above the equality limit satisfying the inequality relations.\\
\\ In conclusion, our results of single heavy flavour baryons are found to be in
accordance with predictions of more realistic computations like the lattice results
\cite{Bowler1998} and Faddeev approach \cite{Garcilazo2007} as well as with other model
predictions. The behavior of hyperfine mass spiltting studied here with respect to the
regularization parameter, $r_{0}$ and the resultant saturation property of the mass
spiltting as seen in Fig \ref{fig:01} and consequently for the masses of $\Lambda_{Q}$
(Q$\in$b,c) seen in Fig \ref{fig:02} and 3 in the range of $r_{0}\leq 10^{-4}$
Me$V^{-1}$ is an important feature that might help us to understand the nature of gluon
exchange interactions between the pair of quarks inside the Baryon. For $\Lambda_{Q}$
Baryons, the extra 200 Me$V$, at the saturation range of $r_{0}\leq 10^{-4}$
Me$V^{-1}$, above their experimental masses indicates the requirement of an attractive
interaction potential for the description of $\Lambda_{Q}$ Baryons which is independent
of the potential model index as well as the heavy flavour content. Detail analysis of
these issues related to the single heavy flavour baryons will be more meaningful when
we acquire more experimental information about these states and their orbital
excitations. Some of
these baryons and orbital excitation are expected to be observed at future experiments.\\
\\
It is important to note that the predictions of the magnetic moment of single heavy
Baryons studied here are with no additional parameters. Our results on magnetic moments
are compared with one of our recent predictions using hypercentral potential
\cite{Bhavin2008} and predictions based on Faddeev formalism for the baryons
\cite{B.Silvestre-brac1996} as well as with relativistic (RQM) and the nonrelativistic
quark model (NRQM) predictions of \cite{Amand2006} in Table \ref{tab:06}. The special
feature of the present study to compute the magnetic moments of single heavy flavour
baryons is the consideration of the effective interactions of the bound state quarks by
defining an effective bound state mass to the quarks within the baryon, which vary
according to different interquark potential as well as with quark compositions.\\
\\
Experimental measurements of the heavy flavour baryon magnetic moments are sparse and
only few experimental groups (BTeV and SELEX Collaborations) are expected to do
measurements in near future. \\
\\
We conclude that the interquark potential interactions and the particular spin
hyperfine structure assumed in the present study play significant role in the
description of heavy flavour baryonic properties in particular for their spin hyperfine
splitting and magnetic moments. We look forward future observations of more
experimental baryonic states in the single heavy flavour sector at different heavy
flavour high luminosity experiments.

\section{Acknowledgments}
We are extremely thankful to the referees for their valuable suggestions. We
acknowledge the financial support from University Grant Commission, Government of
India, under a Major Research Project \textbf{F. 32-31/2006(SR)}.\\

\end{document}